\pdfoutput=1
\RequirePackage[T1]{fontenc}
\documentclass[11pt]{article}
\usepackage[skip=1em plus 0.2em minus 0.1em]{parskip}
\usepackage[height=8.85in,width=6.45in]{geometry}

\usepackage[font=small]{caption}
\usepackage{subcaption}
\usepackage[utf8]{inputenc}
\usepackage{amsmath}
\usepackage{amssymb}
\usepackage{mathtools}
\numberwithin{equation}{section}
\usepackage{slashed}
\usepackage{braket}
\usepackage{enumitem}

\usepackage{newtxtext} %
\usepackage{setspace}            %
\usepackage{titlesec}            %
\usepackage{xspace}
\usepackage{graphicx}
\usepackage[most]{tcolorbox}
\usepackage{ragged2e}

\usepackage[svgnames]{xcolor}
\usepackage[colorlinks,citecolor=DarkGreen,linkcolor=FireBrick,urlcolor=FireBrick,linktocpage,unicode]{hyperref}
\usepackage{nicematrix}
\usepackage{nicefrac}       %
\usepackage[bottom]{footmisc}
\usepackage{array} %
\usepackage{adjustbox} %
\usepackage{longtable, lscape}

\makeatletter
\@ifundefined{PRISM@MACROS@LOADED}{\gdef\PRISM@MACROS@LOADED{}}{ }
\makeatother

\usepackage{graphicx}
\usepackage{tikz}
\usepackage{tikz-cd}
\usepackage{bm}
\usepackage{physics}
\usepackage{xcolor}
\usepackage{natbib}
\usepackage{mdframed}
\usepackage{nicefrac}
\usepackage{booktabs}
\usepackage{lipsum}
\usepackage{titlesec}
\usepackage{wrapfig,lipsum,booktabs}
\usepackage{blindtext}
\usepackage{algorithm}
\usepackage{algorithmic}
\usepackage{titletoc}
\usepackage{todonotes}
\usepackage{titletoc}
\usepackage{setspace}
\usepackage{dsfont}
\usepackage{adjustbox}
\usepackage{caption}
\usepackage{subcaption}
\usepackage{longtable, lscape}

\usepackage[capitalize,noabbrev]{cleveref}

\usepackage{CJK}

\usepackage{array}
\usepackage{makecell}

\newcommand{\bea}{\begin{eqnarray}}
\newcommand{\eea}{\end{eqnarray}}

\usepackage{slashed}

\def\({\left(}
\def\){\right)}
\def\[{\left[}
\def\]{\right]}

\usepackage{dashbox}
\usepackage{xcolor}
\usepackage{colortbl}
\definecolor{lightyellow}{rgb}{1.0, 0.95, 0.7}
\definecolor{Blue}{rgb}{0, 0, 0.8}
\definecolor{blue}{rgb}{0,0,1}
\definecolor{darkgreen}{rgb}{0,0.40,0}
\definecolor{firebrick}{rgb}{0.698,0.133,0.133}

\definecolor{colorA}{rgb}{1,0,0}
\definecolor{colorB}{rgb}{0,0.3,1}
\definecolor{colorC}{rgb}{0.9,0.8,0.2}
\definecolor{colorD}{rgb}{0,0.65,0}
\definecolor{lesslightgray}{rgb}{0.5,0.5,0.5}
\definecolor{light-gray}{gray}{0.95}

\let\cite\citep

\usepackage{amsthm}
\makeatletter
\def\th@remark{%
  \thm@headfont{\bfseries}%
  \normalfont %
  \thm@preskip\topsep \divide\thm@preskip\tw@
  \thm@postskip\thm@preskip
}
\makeatother
\theoremstyle{definition}

\tcolorboxenvironment{theorem}{
  breakable,
  colback=black!10,
  colframe=white,%
  width=\dimexpr\linewidth+10pt\relax,%
  enlarge left by=-5pt,%
  enlarge right by=-5pt,%
  boxsep=5pt,%
  boxrule=0pt,
  left=0pt,right=0pt,top=0pt,bottom=0pt,
  sharp corners,
  before skip=\topsep,
  after skip=\topsep
}

\tcolorboxenvironment{lemma}{
  breakable,
  colback=black!10,
  colframe=white,%
  width=\dimexpr\linewidth+10pt\relax,%
  enlarge left by=-5pt,%
  enlarge right by=-5pt,%
  boxsep=5pt,%
  boxrule=0pt,
  left=0pt,right=0pt,top=0pt,bottom=0pt,
  sharp corners,
  before skip=\topsep,
  after skip=\topsep
}

\tcolorboxenvironment{corollary}{
  breakable,
  colback=black!10,
  colframe=white,%
  width=\dimexpr\linewidth+10pt\relax,%
  enlarge left by=-5pt,%
  enlarge right by=-5pt,%
  boxsep=5pt,%
  boxrule=0pt,
  left=0pt,right=0pt,top=0pt,bottom=0pt,
  sharp corners,
  before skip=\topsep,
  after skip=\topsep
}

\theoremstyle{definition}

\tcolorboxenvironment{definition}{
  breakable,
  colback=black!10,
  colframe=white,%
  width=\dimexpr\linewidth+10pt\relax,%
  enlarge left by=-5pt,%
  enlarge right by=-5pt,%
  boxsep=5pt,%
  boxrule=0pt,
  left=0pt,right=0pt,top=0pt,bottom=0pt,
  sharp corners,
  before skip=\topsep,
  after skip=\topsep
}
\theoremstyle{remark}

\tcolorboxenvironment{assumption}{
  breakable,
  colback=black!10,
  colframe=white,%
  width=\dimexpr\linewidth+10pt\relax,%
  enlarge left by=-5pt,%
  enlarge right by=-5pt,%
  boxsep=5pt,%
  boxrule=0pt,
  left=0pt,right=0pt,top=0pt,bottom=0pt,
  sharp corners,
  before skip=\topsep,
  after skip=\topsep
}

\tcolorboxenvironment{problem}{
  breakable,
  colback=black!10,
  colframe=white,%
  width=\dimexpr\linewidth+10pt\relax,%
  enlarge left by=-5pt,%
  enlarge right by=-5pt,%
  boxsep=5pt,%
  boxrule=0pt,
  left=0pt,right=0pt,top=0pt,bottom=0pt,
  sharp corners,
  before skip=\topsep,
  after skip=\topsep
}

\crefname{theorem}{Theorem}{Theorems}
\crefname{proposition}{Proposition}{Propositions}
\crefname{lemma}{Lemma}{Lemmas}
\crefname{corollary}{Corollary}{Corollaries}
\crefname{definition}{Definition}{Definitions}
\crefname{assumption}{Assumption}{Assumptions}
\crefname{remark}{Remark}{Remarks}
\crefname{problem}{Problem}{Problems}
\crefname{property}{Property}{property}

\numberwithin{equation}{section}
\numberwithin{theorem}{section}
\numberwithin{proposition}{section}
\numberwithin{definition}{section}
\numberwithin{lemma}{section}
\numberwithin{assumption}{section}
\numberwithin{remark}{section}

\usepackage{lipsum}

\makeatletter
\@ifundefined{widebar}{%
\let\save@mathaccent\mathaccent
\newcommand*\if@single[3]{%
    \setbox0\hbox{${\mathaccent"0362{#1}}^H$}%
    \setbox2\hbox{${\mathaccent"0362{\kern0pt#1}}^H$}%
    \ifdim\ht0=\ht2 #3\else #2\fi
}
\newcommand*\rel@kern[1]{\kern#1\dimexpr\macc@kerna}
\newcommand*\widebar[1]{\@ifnextchar^{{\wide@bar{#1}{0}}}{\wide@bar{#1}{1}}}
\newcommand*\wide@bar[2]{\if@single{#1}{\wide@bar@{#1}{#2}{1}}{\wide@bar@{#1}{#2}{2}}}
\newcommand*\wide@bar@[3]{%
    \begingroup
    \def\mathaccent##1##2{%
        \let\mathaccent\save@mathaccent
        \if#32 \let\macc@nucleus\first@char \fi
        \setbox\z@\hbox{$\macc@style{\macc@nucleus}_{}$}%
        \setbox\tw@\hbox{$\macc@style{\macc@nucleus}{}_{}$}%
        \dimen@\wd\tw@
        \advance\dimen@-\wd\z@
        \divide\dimen@ 3
        \@tempdima\wd\tw@
        \advance\@tempdima-\scriptspace
        \divide\@tempdima 10
        \advance\dimen@-\@tempdima
        \ifdim\dimen@>\z@ \dimen@0pt\fi
        \rel@kern{0.6}\kern-\dimen@
        \if#31
        \overline{\rel@kern{-0.6}\kern\dimen@\macc@nucleus\rel@kern{0.4}\kern\dimen@}%
        \advance\dimen@0.4\dimexpr\macc@kerna
        \let\final@kern#2%
        \ifdim\dimen@<\z@ \let\final@kern1\fi
        \if\final@kern1 \kern-\dimen@\fi
        \else
        \overline{\rel@kern{-0.6}\kern\dimen@#1}%
        \fi
    }%
    \macc@depth\@ne
    \let\math@bgroup\@empty \let\math@egroup\macc@set@skewchar
    \mathsurround\z@ \frozen@everymath{\mathgroup\macc@group\relax}%
    \macc@set@skewchar\relax
    \let\mathaccentV\macc@nested@a
    \if#31
    \macc@nested@a\relax111{#1}%
    \else
    \def\gobble@till@marker##1\endmarker{}%
    \futurelet\first@char\gobble@till@marker#1\endmarker
    \ifcat\noexpand\first@char A\else
    \def\first@char{}%
    \fi
    \macc@nested@a\relax111{\first@char}%
    \fi
    \endgroup
    }
}{}
\makeatother

\makeatletter
\newcommand*{\redefinesymbolwitharg}[1]{%
  \expandafter\let\csname ltx#1\expandafter\endcsname\csname #1\endcsname
  \@namedef{#1}{\@ifnextchar{^}{\@nameuse{#1@}}{\@nameuse{#1@}^{}}}%
  \expandafter\def\csname #1@\endcsname^##1##2{%
     \csname ltx#1\endcsname\ifx!##1!\else^{##1}\fi\mathopen{}\mathclose\bgroup\left(##2\aftergroup\egroup\right)
     }%
}
\makeatother
\redefinesymbolwitharg{sin}
\redefinesymbolwitharg{cos}

\usepackage{tocloft}

\setlist[itemize,enumerate]{
  parsep=\parskip,                                   %
  itemsep=\dimexpr .3em - \parskip\relax plus 2pt,   %
  topsep=\dimexpr 6pt - \parskip\relax plus 1pt minus 1pt,
  partopsep=0pt,
  listparindent=\parindent
}

\newif\ifrevision
\revisiontrue  %

\newcolumntype{C}[1]{>{\collectcell\cellcolorcontents{LightCyan}}c<{#1}} %

\usepackage{multirow}

\begin{document}

\thispagestyle{empty}

\noindent\begin{tcolorbox}[
  enhanced,
  frame hidden,
  colback=Lavender!50,
  colframe=Lavender!50,
  arc=12pt,
  left=16pt,right=16pt,top=16pt,bottom=16pt,
  width=\textwidth
]
\raggedright

{\LARGE\sffamily\bfseries Cell-JEPA: Latent Representation Learning for Single-Cell Transcriptomics
\par}

\vspace{0.8em}

{\normalsize\sffamily
\textbf{Ali ElSheikh}$^{*,1}$,
\textbf{Rui-Xi Wang}$^{*,3,6}$,
\textbf{Weimin Wu}$^{*,1}$, 
\textbf{Yibo Wen}$^{1}$, 
\textbf{Payam Dibaeinia}$^{2}$,
\textbf{Jennifer Yuntong Zhang}$^{5}$,
\textbf{Jerry Yao-Chieh Hu}$^{1}$,
\textbf{Mei Knudson}$^{2,4}$,
\textbf{Sudarshan Babu}$^{2}$,
\textbf{Shao-Hua Sun}$^{6}$,
\textbf{Aly A. Khan}$^{2,4}$,
\textbf{Han Liu}$^{\dagger,1}$\\
\par}

\vspace{0.35em}

{\normalsize
$^{1}$Northwestern University, $^{2}$Biohub, $^{3}$Massachusetts Institute of Technology, $^{4}$University of Chicago, $^{5}$University of Toronto, $^{6}$National Taiwan University\par}

\vspace{1.0em}

{\justifying
\textbf{Abstract:} Single-cell foundation models learn by reconstructing masked gene expression, 
implicitly treating technical noise as signal. With dropout rates exceeding 
90\%, reconstruction objectives encourage models to encode measurement artifacts 
rather than stable cellular programs. We introduce Cell-JEPA, a joint-embedding 
predictive architecture that shifts learning from reconstructing sparse counts 
to predicting in latent space. The key insight is that cell identity is 
redundantly encoded across genes. We show predicting cell-level embeddings from 
partial observations forces the model to learn dropout-robust features. On 
cell-type clustering, Cell-JEPA achieves 0.72 AvgBIO in zero-shot transfer 
versus 0.53 for scGPT, a 36\% relative improvement. On perturbation prediction 
within a single cell line, Cell-JEPA improves absolute-state reconstruction but 
not effect-size estimation, suggesting that representation learning and 
perturbation modeling address complementary aspects of cellular prediction.
\par}

\vspace{0.8em}

\par

\end{tcolorbox}

\begingroup
\renewcommand{\thefootnote}{\fnsymbol{footnote}}\setcounter{footnote}{0}
\footnotetext[1]{\textit{Equal contribution.}}
\footnotetext[2]{Correspondence: Han Liu (\texttt{hanliu@northwestern.edu}).}
\endgroup

\setcounter{footnote}{0}

\begin{figure*}[htbp]
    \centering
    \includegraphics[width=\textwidth]{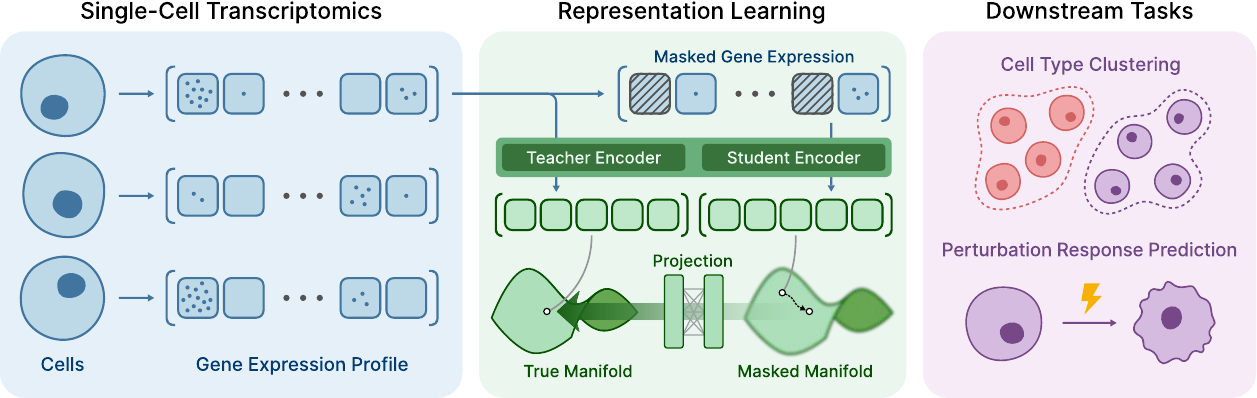}
    \caption{
        \textbf{Overview of the Cell-JEPA Pipeline.}
Raw biological cells undergo single-cell RNA sequencing to generate high-dimensional, sparse gene expression profiles.
The profiles are passed through a joint embedding architecture, where a student encoder receives masked inputs and predicts the stable latent representations produced by a teacher encoder from unmasked data. This predictive alignment learns a robust biological manifold that enables downstream tasks, including cell-type clustering and perturbation response prediction.
    }   
    \label{fig:overview}
    \vspace{-0.5em}
\end{figure*}

\clearpage
\section{Introduction}
\label{sec:intro}

We introduce Cell-JEPA, a self-supervised foundation model for single-cell transcriptomics. 
Single-cell transcriptomics enables the study of cellular heterogeneity, development, and disease \cite{Ding2022, Svensson2018, Tang2011}.
Each cell is represented by an expression profile across tens of thousands of genes, providing a high-dimensional snapshot of cellular state.
However, single-cell measurements are sparse and noisy due to limited molecular capture, stochastic sampling, and experimental variability \cite{Luecken2019-mg}.
Therefore, the core challenge is to infer stable and biologically meaningful cellular programs from high-dimensional and noisy observations.

Recent progress has introduced large-scale self-supervised foundation models for single-cell data.
A common strategy is to learn representations by predicting masked or noisy parts of the transcriptome.
For example, scGPT \cite{Cui2024} performs generative pre-training by predicting expression values for a subset of ``unknown'' genes given observed ``known'' genes, using a specialized attention mask and an iterative generation procedure that progressively incorporates high-confidence predictions.
Geneformer \cite{Theodoris2023} adopts a BERT-style masked learning objective on ranked gene tokens, training to recover masked genes from the remaining context within each cell.
UCE \cite{Rosen2023} relies on self-supervision by masking expressed genes and training to predict gene-level expression presence, while aiming to provide a transferable embedding space across diverse tissues and species.
In parallel, scVI \cite{Lopez2018} learns low-dimensional cell states through a probabilistic generative model of counts trained by variational inference.
While these approaches differ in modeling assumptions, they share the key property that pre-training is strongly tied to reconstructing information closely aligned with the observed measurement space (e.g., masked counts, masked gene identities, or gene-level presence/absence).

A limitation of this paradigm is that single-cell expression values can be heavily influenced by sequencing depth, capture efficiency, and technical noise.
As a result, objectives that emphasize accurate gene-level reconstruction may encourage models to represent measurement-specific variability, instead of prioritizing stable structure that defines cellular identity and regulatory programs.
This motivates learning objectives that constrain the \emph{representation space}, rather than optimizing for fidelity to noisy observed entries.

Recent progress in computer vision suggests that \emph{Joint Embedding Predictive Architectures} (JEPA) provide a strong self-supervised signal by predicting masked content in a learned latent space, rather than reconstructing raw pixel inputs~\cite{assran2025vjepa2selfsupervisedvideo,  assran2023selfsupervisedlearningimagesjointembedding, lecun_2022}.
By shifting prediction to the embedding space, JEPA encourages consistency at the level of abstract features and reduces reliance on exact input values, which is desirable when observations are noisy and incomplete.
This property closely matches the goal of single-cell modeling: recovering robust cell states from sparse and noisy gene expression measurements.

Motivated by these developments, we introduce \textbf{Cell-JEPA}, a JEPA-style framework for single-cell representation learning built on top of scGPT.
Cell-JEPA augments reconstruction-style pre-training with a latent-space prediction objective that enforces consistency between predicted and target cell representations under partial observation.
We hypothesize that combining scGPT’s scalable transcriptome modeling with JEPA-style latent prediction yields embeddings that better capture stable cellular programs and higher-order biological structure.
Across downstream tasks, we show that Cell-JEPA learns more robust representations, improving performance on cell-state prediction problems compared to reconstruction-only baselines. 

In summary, we have the following three key contributions.
\begin{itemize}
    \item We propose \textbf{Cell-JEPA}, a predictive self-supervised framework for single-cell transcriptomics that complements reconstruction-based pre-training with latent-space prediction.
    It enables stable and meaningful cellular representation learning from high-dimensional, sparse, and noisy gene expression.
    
    \item We curate a large-scale human kidney scRNA-seq corpus comprising 800,000 cells spanning multiple studies, donors, and experimental conditions.
    Based on this, we pre-train and release our single-cell foundation model under this new pre-training paradigm.
    
    \item We evaluate Cell-JEPA in three downstream settings: (i) supervised finetuning for cell-type clustering, (ii) zero-shot transfer for cell-type clustering, and (iii) perturbation-response prediction.
    Across all experiments, Cell-JEPA outperforms the baseline scGPT model.
    In particular, Cell-JEPA achieves a relative improvement of $35\%$ on the primary evaluation metric in the zero-shot setting.
\end{itemize}

\paragraph{Organization.} 
We describe the pre-training and finetuning methods in \cref{sec:method}, present experimental results in \cref{sec:exp}, and review related work in \cref{sec:related}.

\section{Method}
\label{sec:method}
\begin{figure*}[t]
    \centering
    \includegraphics[width=\textwidth]{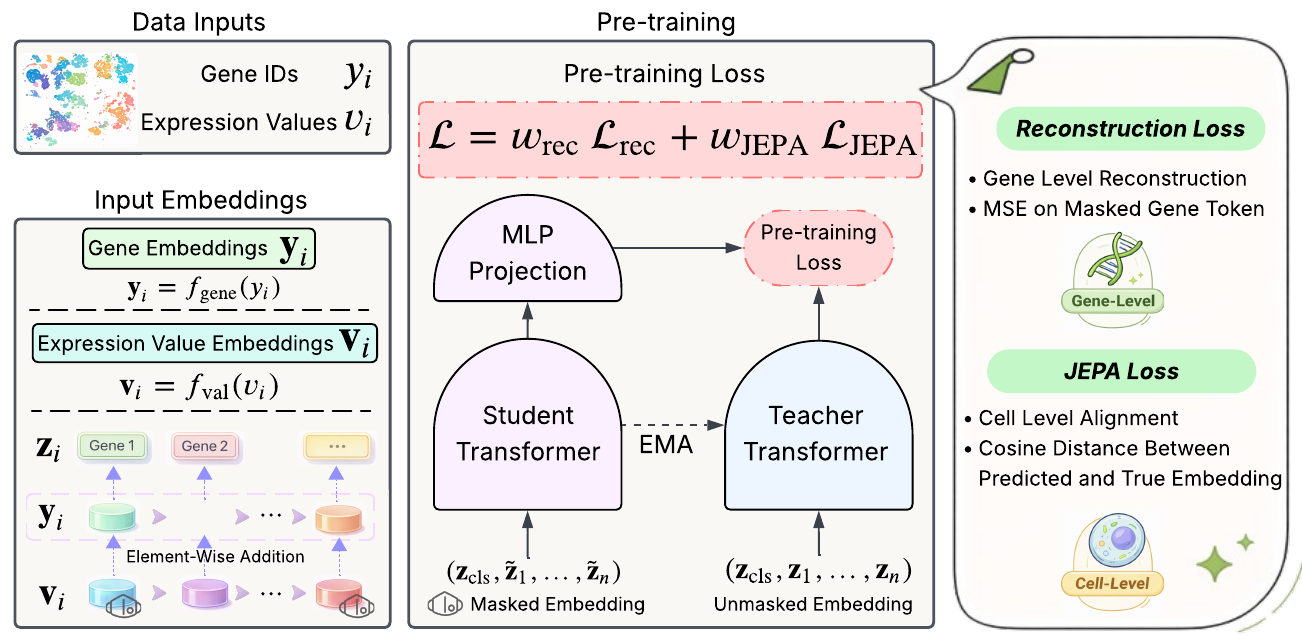}
    \caption{
        \textbf{Cell-JEPA Model Architecture and Training Pipeline.}
Cell-JEPA builds on scGPT with a student–teacher transformer architecture.
Gene identities and discretized expression values are embedded and summed to form input token embeddings.
The student encoder processes masked inputs, while the teacher encoder processes the corresponding unmasked inputs and is updated via an exponential moving average (EMA).
Training optimizes a gene-level reconstruction loss in expression space and a JEPA loss that predicts the teacher's cell embedding from the masked student view.
    }   
    \label{fig:cell_jepa_pipeline}
\end{figure*}

In this section, we first describe the data preprocessing used for Cell-JEPA pre-training in \cref{subsec:pretrain_data_process}.
We then present the tokenization procedure and model architecture in \cref{sec:arch}.
Next, we detail the pre-training objective in \cref{sec:pretrain_loss}.
Finally, we describe the downstream finetuning losses in \cref{sec:finetune-loss}.

\paragraph{Dataset Formatting.}
Each scRNA-seq dataset is represented as a sparse cell-by-gene count matrix. Following prior work \citep{Cui2024}, we exclude genes with zero counts within each cell and store the remaining non-zero entries as aligned lists of gene identifiers and expression values. 
We encode each cell as a sparse sequence $\{(y_k, v_k)\}_{k=1}^{L}$, where $y_k$ denotes a gene identity and $v_k \in \mathbb{R}_{+}$ is its corresponding expression value.

\subsection{Pre-training Dataset Processing} 
\label{subsec:pretrain_data_process}

We follow the preprocessing strategy of scGPT \citep{Cui2024} to (i) discretize expression values via per-cell quantile binning and (ii) subsample genes to control sequence length across cells.

\paragraph{Quantile-based Binning.}
Following scGPT \citep{Cui2024}, we discretize non-zero gene expression values using per-cell quantile binning to improve robustness to scale differences across cells.
For each cell, we compute quantile bin edges over its non-zero expression values and map each value to a bin index in $\{1,\dots,B\}$, with $B=50$ by default.
These bin indices serve as the value inputs $v_k$, preserving the relative ordering of expression levels within a cell while reducing dependence on raw count magnitudes.

\paragraph{Gene Subsampling.}
Cells express variable numbers of genes, resulting in sequences with different lengths.
Following scGPT \citep{Cui2024}, we cap the number of expressed genes per cell at $L_{\max}$ (set to $L_{\max}{=}600$).
For cells exceeding this limit, we avoid selecting a fixed subset of genes (e.g., the first $L_{\max}$), which could introduce bias.
Instead, we uniformly subsample $L_{\max}$ gene tokens at random from the expressed gene set, ensuring that each expressed gene has equal probability of being included. This stochastic subsampling also serves as a form of biological dropout: across training epochs, the model observes different partial views of the same cell and must learn from incomplete information.
As a result, the learned representations become more robust to sparsity and expression variability.

\subsection{Tokenization and Model Architecture}
\label{sec:arch}

\paragraph{Overview.}
Cell-JEPA builds directly on scGPT \citep{Cui2024}.
scGPT is a bidirectional Transformer encoder trained with a gene expression prediction objective under masked inputs (see \cref{sec:scgpt}).
We adopt the same tokenization, embedding design, and encoder backbone as scGPT, and introduce a Joint Embedding Predictive Architecture (JEPA) by instantiating two copies of the scGPT encoder: a \emph{student} network optimized by gradient descent and a slowly-evolving \emph{teacher} network updated via exponential moving average (EMA).
This student--teacher formulation enables us to add a representation-level JEPA objective alongside the original scGPT reconstruction-style loss.
We detail the tokenization procedure, the teacher–student model design, and the encoder architecture as follows.

\subsubsection{Tokenization and Input Embeddings}
\paragraph{Vocabulary and Special Tokens.}
Following scGPT \citep{Cui2024}, we construct a unified gene vocabulary and augment it with special tokens including \texttt{<cls>} and \texttt{<pad>}.
For each cell, we prepend a \texttt{<cls>} token, and pad sequences within a minibatch to length $L_{\max}+1$ using \texttt{<pad>}.

\paragraph{Gene and Value Embeddings.}
Given a gene identity $y_i$ and its (preprocessed) expression value $v_i$, we embed the gene token and value using the same design as scGPT:
\begin{align*}
\mathbf{y}_i = f_{\mathrm{gene}}(y_i), 
\quad
\mathbf{v}_i = f_{\mathrm{val}}(v_i), \notag
\end{align*}
where $f_{\mathrm{gene}}$ is a learnable embedding lookup table and $f_{\mathrm{val}}$ is a small MLP mapping scalar inputs to the model hidden dimension.
The input representation for gene $i$ is the sum:
\begin{align}
\mathbf{z}_i = \mathbf{y}_i + \mathbf{v}_i.
\label{eq:token_embed_sum}
\end{align}
The \texttt{<cls>} token embedding is obtained from the same gene embedding table:
\begin{align*}
\mathbf{z}_{\mathrm{cls}} = f_{\mathrm{gene}}(\texttt{<cls>}).
\end{align*}

\paragraph{Masking.}
During pre-training, we apply masking to \emph{expression values only}, following scGPT \citep{Cui2024}.
For a masked gene $i$, we replace $v_i$ with a sentinel value $v_{\mathrm{mask}}$ (set to $-1$), producing a masked value embedding:
\begin{align*}
\mathbf{v}_{\mathrm{mask}} = f_{\mathrm{val}}(v_{\mathrm{mask}}),
\end{align*}
and a masked token representation:
\begin{align*}
\tilde{\mathbf{z}}_i = \mathbf{y}_i + \mathbf{v}_{\mathrm{mask}}.
\end{align*}
Let $\mathbf{Z} = (\mathbf{z}_{\mathrm{cls}}, \mathbf{z}_1,\ldots,\mathbf{z}_n)$ denote the unmasked token sequence and $\tilde{\mathbf{Z}} = (\mathbf{z}_{\mathrm{cls}}, \tilde{\mathbf{z}}_1,\ldots,\tilde{\mathbf{z}}_n)$ the corresponding masked sequence.

\subsubsection{Student-teacher Encoders}
\paragraph{Encoders.}
 Both the student encoder $g_{\mathrm{S}}$ and teacher encoder $g_{\mathrm{T}}$ share the \emph{same architecture as scGPT} \citep{Cui2024} (i.e., a bidirectional Transformer encoder stack).
The student processes $\mathbf{Z}$, while the teacher processes $\tilde{\mathbf{Z}}$ to produce stable target representations:
\begin{align*}
\hat{\mathbf{H}} = g_{\mathrm{S}}(\tilde{\mathbf{Z}}), 
\quad
\mathbf{H} = g_{\mathrm{T}}(\mathbf{Z}),
\end{align*}
We use the \texttt{<cls>} outputs as cell-level embeddings:
\begin{align}
\hat{\mathbf{e}} = \hat{\mathbf{H}}_{0}, \quad \mathbf{e} = \mathbf{H}_{0}.
\label{eq:cls_embeddings}
\end{align}

\paragraph{EMA Teacher Update.}
The teacher parameters are updated as an exponential moving average of the student parameters (as commonly used in JEPA-style methods):
\begin{align*}
\theta_{\mathrm{T}} \leftarrow m\,\theta_{\mathrm{T}} + (1-m)\,\theta_{\mathrm{S}},
\end{align*}
where $m\in[0,1)$ is the momentum coefficient, and $\theta_{\mathrm{S}},\theta_{\mathrm{T}}$ are the parameters of the student and teacher encoders, respectively.
This yields a slowly-evolving target encoder that stabilizes the representation prediction objective.

\subsubsection{Transformer Block and Attention Masking}
Both encoders use the same transformer block structure as scGPT \citep{Cui2024}.
Each block consists of self-attention layers, multilayer perceptrons, nonlinear transformations, and layer normalization.
We describe the self-attention layer in detail below for clarity.

\paragraph{Self-attention.}
For a hidden sequence $\mathbf{H}\in\mathbb{R}^{(L+1)\times D}$, where $D$ is the embedding dimension.
We use $\mathbf{A}_{\mathrm{mask}}$ to denote a structured attention mask to prevent information leakage between masked targets.
Let $\mathcal{U}_{\mathrm{mask}}$ denote the indices of masked genes at the current step, and then the corresponding entries in $\mathbf{A}_{\mathrm{mask}}$ are:
\begin{align*}
a_{i,j} =
\begin{cases}
0, & \text{if } j \notin \mathcal{U}_{\mathrm{mask}},\\
0, & \text{if } i=j \text{ and } j\in \mathcal{U}_{\mathrm{mask}},\\
-\infty, & \text{otherwise}.
\end{cases}
\end{align*}
This design ensures that each masked token can attend to all unmasked tokens (and itself), but not to other masked tokens.
We then compute the attention module as:
\begin{align*}
&~ \mathrm{Attn}(\mathbf{H}) = 
\mathrm{Softmax}\left(\frac{\mathbf{Q}\mathbf{K}^{\top}}{\sqrt{d}} + \mathbf{A}_{\mathrm{mask}} \right)\mathbf{V},
\end{align*}
where $\mathbf{Q} = \mathbf{H}\mathbf{W}_Q$, $
\mathbf{K} = \mathbf{H}\mathbf{W}_K$, $
\mathbf{V} = \mathbf{H}\mathbf{W}_V$ and $\mathbf{W}_Q,\mathbf{W}_K,\mathbf{W}_V\in\mathbb{R}^{D\times d}$ are the projection matrices, and $d$ denotes the attention head dimension.

\subsection{Pre-training Loss Calculation}
\label{sec:pretrain_loss}

Cell-JEPA augments the original scGPT pre-training objective \citep{Cui2024} with a JEPA-style representation prediction loss.
Concretely, we optimize the student encoder using (i) a gene-level reconstruction loss identical in spirit to scGPT and (ii) a cell-level JEPA objective that predicts the teacher’s latent embedding from the masked student view.

\paragraph{JEPA Objective.}
Let $\hat{\mathbf{e}}$ and $\mathbf{e}$ denote the student and teacher \texttt{<cls>} embeddings from \eqref{eq:cls_embeddings}.
We introduce a predictor head $p(\cdot)$ and apply it to the student embedding:
\begin{align*}
    \tilde{\mathbf{e}} = p(\hat{\mathbf{e}}).
\end{align*}
We then minimize a cosine-distance loss between the predicted student embedding and the teacher target embedding:
\begin{align}
\mathcal{L}_{\mathrm{JEPA}}
= 1 - \frac{\tilde{\mathbf{e}}^{\top}\,\mathrm{sg}(\mathbf{e})}
{\|\tilde{\mathbf{e}}\|_2\,\|\mathrm{sg}(\mathbf{e})\|_2},
\label{eq:JEPA_Loss}
\end{align}
where $\mathrm{sg}(\cdot)$ denotes stop-gradient (the teacher embedding is treated as a fixed target).
Unlike reconstruction losses defined in expression space, $\mathcal{L}_{\mathrm{JEPA}}$ operates purely in representation space and encourages the student to match the teacher’s cell-level semantics under masking.

\paragraph{Gene-level Reconstruction Loss.}
In addition to $\mathcal{L}_{\mathrm{JEPA}}$, we retain the scGPT-style gene expression prediction objective~\citep{Cui2024}. 
Unlike scGPT’s iterative generation procedure, we predict \emph{all} masked genes in a single forward pass, following a BERT-style masked prediction scheme.

Given the student token outputs $\hat{\mathbf{H}} = g_{\mathrm{S}}(\tilde{\mathbf{Z}})$, we predict expression values for each token using a value head $r(\cdot)$ (MLP applied row-wise):
\begin{align*}
\hat{\mathbf{v}} = r(\hat{\mathbf{H}}),
\end{align*}
where $\hat{\mathbf{v}}_i$ denotes the reconstructed value for gene token $i$.
We compute mean squared error on the masked gene set $\mathcal{U}_{\mathrm{mask}}$:
\begin{align}
\mathcal{L}_{\mathrm{rec}}
=
\frac{1}{|\mathcal{U}_{\mathrm{mask}}|}
\sum_{i \in \mathcal{U}_{\mathrm{mask}}}
\left(\hat{\mathbf{v}}_i - v_i\right)^2.
\label{eq:pre-train_loss}
\end{align}
This term preserves scGPT’s training signal by explicitly encouraging accurate prediction of masked expression values from the observed context.

\paragraph{Combined Pre-training Objective.}
The final pre-training objective is a weighted combination of reconstruction and JEPA losses:
\begin{align*}
\mathcal{L}_{\mathrm{pre-train}}
=
w_{\mathrm{rec}}\,\mathcal{L}_{\mathrm{rec}}
+
w_{\mathrm{JEPA}}\,\mathcal{L}_{\mathrm{JEPA}},
\end{align*}
where $w_{\mathrm{rec}}$ and $w_{\mathrm{JEPA}}$ control the relative contribution of each term.
Intuitively, $\mathcal{L}_{\mathrm{rec}}$ anchors learning to gene-level expression signals as in scGPT, while $\mathcal{L}_{\mathrm{JEPA}}$ enforces consistency of cell-level representations in latent space.
Optimizing both jointly yields embeddings that retain fine-grained information for expression recovery while improving robustness and semantic structure through representation-level prediction.

\subsection{Downstream Finetuning}
\label{sec:finetune-loss}
During downstream finetuning, we optimize Cell-JEPA using four objectives:
(i) gene expression prediction (GEP);
(ii) gene expression prediction for cell modeling (GEPC);
(iii) elastic cell similarity (ECS); and
(iv) JEPA prediction loss.
We adopt GEP, GEPC, and ECS from scGPT \citep{Cui2024}, and extend the finetuning objective by additionally retaining $\mathcal{L}_{\mathrm{JEPA}}$.

\subsubsection{GEP and GEPC Loss}
\paragraph{Gene Expression Prediction (GEP).} 
GEP is equivalent to the reconstruction loss $\mathcal{L}_{rec}$ in \eqref{eq:pre-train_loss} used during pre-training. 
Using the student predictions $\hat{\mathbf{v}}_i$, we minimize mean squared error over the masked gene set~$\mathcal{U}_{\mathrm{mask}}$:
\begin{align*}
    \mathcal{L}_\mathrm{GEP} = \frac{1}{|\mathcal{U}_{\mathrm{mask}}|}
\sum_{i \in \mathcal{U}_{\mathrm{mask}}}
\left(\hat{\mathbf{v}}_i - v_i\right)^2.
\end{align*}
\paragraph{Gene Expression Prediction for Cell Modeling (GEPC).} GEPC is a complementary objective from scGPT to explicitly tie gene-level prediction to the global cell representation \citep{Cui2024}.
For each gene token $i$, we compute a query vector from its gene embedding $\mathbf{y}_i$ and predict expression by an inner product with the cell embedding $\hat{\mathbf{e}}$:
\begin{align*} 
\tilde{\mathbf{v}}_i = f(\mathbf{y}_i)^T \mathbf{W} \hat{\mathbf{e}}, 
\end{align*} 
where $f(\cdot)$ is an MLP and $\mathbf{W}$ is a learnable weight matrix. 
The GEPC loss is the MSE over masked genes: %
\begin{align*}
    \mathcal{L}_\mathrm{GEPC} = \frac{1}{|\mathcal{U}_\mathrm{mask}|}\sum_{i\in\mathcal{U}_\mathrm{mask}}(\tilde{\mathbf{v}}_i-v_i)^2.
\end{align*}

Together, GEP and GEPC provide complementary supervision: GEP enforces accurate masked-value recovery, while GEPC encourages the global cell embedding to be predictive of gene-level expression.

\subsubsection{ECS Loss}
\label{sec:ECS}
\paragraph{Elastic Cell Similarity (ECS).} ECS regularizes the cell embedding space by encouraging representations of biologically similar cells to be closer.
While scGPT proposes an ECS regularizer based on pairwise similarity thresholds \citep{Cui2024}, we modify this component into the following forms given the intuition described in \cref{sec:scgpt-ecs}.
Consider a minibatch of $N$ cells with embeddings $\{(\mathbf{e}_i, c_i)\}_{i=1}^{N}$, where $\mathbf{e}_i$ represents the embedding of cell $i$ (as defined in \eqref{eq:cls_embeddings}) and $c_i$ is its corresponding cell-type label.
We define the cosine similarity between cells $i$ and $j$ as:
\begin{align*}
s_{ij} = \frac{\cos{\mathbf{e}^i, \mathbf{e}^j}}{\tau}, 
\end{align*}
where $\tau$ is a temperature hyperparameter.

We treat each cell in the minibatch as an anchor in turn.
For a given anchor cell $i$, we identify cells in the minibatch that share the same cell type as:
\begin{align*}
p_{ij} =
\begin{cases}
\frac{1}{|\mathcal{P}(i)|}, & \text{if } c_j = c_i, \\
0, & \text{otherwise}.
\end{cases}
\end{align*}
The contrastive loss for anchor cell $i$ is then given by:
\begin{align*}
\mathcal{L}_{\mathrm{ECS},i}
=
- \sum_{j=1}^{N}
p_{ij}
\log
\frac{\exp(s_{ij})}
{\sum_{k=1}^{N} \exp(s_{ik})}.
\end{align*}
We calculate the elastic cell similarity loss by averaging over all anchors in the minibatch:
\begin{align*}
\mathcal{L}_{\mathrm{ECS}}
=
\frac{1}{N}
\sum_{i=1}^{N} \mathcal{L}_{\mathrm{ECS},i}.
\end{align*}
This objective acts as a continuous regularizer to enforce structured similarity in the learned cell embedding space.

\subsubsection{JEPA and Overall Loss}
We retain the JEPA loss term $\mathcal{L}_{\mathrm{JEPA}}$ in \eqref{eq:JEPA_Loss} during finetuning.
This allows the model to adapt embedding-space mappings to distributional differences within the dataset.

The overall finetuning objective $\mathcal{L}_{\mathrm{finetune}}$ is defined as a weighted sum of the four losses as follows:
\begin{align*}
\begin{aligned}
        \mathcal{L}_\mathrm{finetune} =  w_\mathrm{GEP}   \mathcal{L}_\mathrm{GEP} + w_\mathrm{GEPC}  \mathcal{L}_\mathrm{{GEPC}}+ w_\mathrm{ECS}  \mathcal{L}_\mathrm{ECS} + w_\mathrm{JEPA}  \mathcal{L}_\mathrm{JEPA}.
\end{aligned}
\end{align*}
For the scGPT baseline, we use the same finetuning objectives excluding $\mathcal{L}_{\mathrm{JEPA}}$:
\begin{align*}
\begin{aligned}
        \mathcal{L}^\mathrm{scGPT}_\mathrm{finetune} = w_\mathrm{GEP}   \mathcal{L}_\mathrm{GEP} + w_\mathrm{GEPC}  \mathcal{L}_\mathrm{{GEPC}} + w_\mathrm{ECS}  \mathcal{L}_\mathrm{ECS}.
\end{aligned}
\end{align*}
\subsection{Perturbation Prediction}

\paragraph{Perturbation Objective.}
We further finetune the pre-trained scGPT and Cell-JEPA models to predict gene perturbation effects. 
To incorporate perturbation information, we embed the perturbation label $p_i$ of gene $i$ using a standard ID embedding layer $f_{\mathrm{perturb}}(\cdot)$ in prediction model:
\begin{align*}
    \mathbf{p}_i = f_{\mathrm{perturb}}(p_i).
\end{align*}

We modify the gene token representation in \eqref{eq:token_embed_sum} by incorporating the perturbation embedding:
\begin{align*}
\tilde{\mathbf{z}}_i^\mathrm{pert}=\mathbf{y}_i+\mathbf{p}_i+\mathbf{v}_i.
\end{align*}
 We pass the perturbed gene embeddings to the student encoder to generate predicted perturbed gene values. Let $\tilde{\mathbf{Z}}_\mathrm{pert} = (\mathbf{z}_\mathrm{cls},\tilde{\mathbf{z}}^\mathrm{pert}_1,...,\tilde{\mathbf{z}}^\mathrm{pert}_n)$, The prediction process is:
\begin{align*}
    \hat{\mathbf{H}}_\mathrm{pert} = g_\mathrm{S}(\tilde{\mathbf{Z}}_\mathrm{pert}), \quad
    \hat{\mathbf{v}}_i^\mathrm{pert} =  r_\mathrm{pert}(\hat{\mathbf{H}}_\mathrm{pert}),
\end{align*}
where $r_\mathrm{pert}$ is an MLP value head. 
We define the perturbation reconstruction loss as:
\begin{align*}
    \mathcal{L}_\mathrm{pert-pred} = \frac{1}{n}\sum_{i=1}^n (\hat{\mathbf{v}}_i^\mathrm{pert} - v_i^\mathrm{pert})^2,
\end{align*}
where $v_i^\mathrm{pert}$ is the ground truth perturbed expression value for gene $i$.
\paragraph{JEPA and Overall Loss.}
Unlike the pre-training and finetuning settings, we do not apply masking to the gene value inputs. We define the student perturbed cell embedding as:
\begin{align*}
    \hat{\mathbf{e}}_\mathrm{pert} = \hat{\mathbf{H}}^\mathrm{pert}_0.
\end{align*}
We construct the target perturbed gene embedding as:
\begin{align*}
    \mathbf{z}_i^\mathrm{pert} = \mathbf{y}_i+\mathbf{p}_i+\mathbf{v}^\mathrm{pert}_i,
\end{align*}
where $\mathbf{v}^\mathrm{pert}_i$ is the perturbed gene expression embedding for gene $i$. Let $\mathbf{Z}_\mathrm{pert} = (\mathbf{z}_\mathrm{cls},\mathbf{z}^\mathrm{pert}_1,...,\mathbf{z}^\mathrm{pert}_n)$.
The target encoder produces:
\begin{align*}
    \mathbf{H}_\mathrm{pert} = g_\mathrm{T}(\mathbf{Z}_\mathrm{pert}), \quad
    \mathbf{e}_\mathrm{pert} = \mathbf{H}^\mathrm{pert}_0.
\end{align*}
We introduce a predictor head $p_\mathrm{pert}(\cdot)$ (an MLP) and apply it to the student perturbed embedding:
\begin{align*}
    \tilde{\mathbf{e}}_\mathrm{pert} = p_\mathrm{pert}(\hat{\mathbf{e}}_\mathrm{pert}).
\end{align*}
We compute the perturbed JEPA loss analogously to \eqref{eq:JEPA_Loss}:
\begin{align*}
\mathcal{L}^\mathrm{pert}_{\mathrm{JEPA}}
= 1 - \frac{\tilde{\mathbf{e}}_\mathrm{pert}^{\top}\,\mathrm{sg}(\mathbf{e}_\mathrm{pert})}
{\|\tilde{\mathbf{e}}_\mathrm{pert}\|_2\,\|\mathrm{sg}(\mathbf{e}_\mathrm{pert})\|_2}.
\end{align*}
We also adopt the ECS formulation from \cref{sec:ECS} to define a perturbed ECS loss, where the similarity between two cells $i$ and $j$ is computed as the cosine similarity between their perturbed cell embeddings $\hat{\mathbf{e}}_i^\mathrm{pert}$ and $\hat{\mathbf{e}}_j^\mathrm{pert}$.

The overall perturbation objective is a weighted sum of all perturbed loss terms:
\begin{align*}
\begin{aligned}
        \mathcal{L}_\mathrm{pert} = w_{\rm pert \text{-} rec}  \mathcal{L}_{\rm pert \text{-} rec} + w_\mathrm{JEPA}^\mathrm{pert} \mathcal{L}_\mathrm{JEPA}^\mathrm{pert}+ w_\mathrm{ECS} \mathcal{L}_\mathrm{ECS}.
\end{aligned}
\end{align*} 
For the scGPT baseline, we use the same objective but exclude the JEPA term:
\begin{align*}
        \mathcal{L}_\mathrm{pert}^\mathrm{scGPT} = w_{\rm pert \text{-} rec} \mathcal{L}_{\rm pert \text{-} rec}
        + w_\mathrm{ECS} \mathcal{L}_\mathrm{ECS}.
\end{align*}

\section{Experimental Studies}
\label{sec:exp}
In this section, we first provide the foundation model pre-training recipe in \cref{subsec:pre_recipe}.
We then evaluate the pre-trained models in three downstream settings: (i) supervised finetuning for cell-type clustering in \cref{subsec:task1}, (ii) zero-shot transfer for cell-type clustering in \cref{subsec:task2}, and (iii) perturbation-response prediction \cref{subsec:perturbation}.

\subsection{Pre-training Recipe}
\label{subsec:pre_recipe}

To validate the benefits of the Cell-JEPA framework, we pre-train two foundation models using the same pre-training dataset and training recipe: Cell-JEPA and scGPT \cite{Cui2024}.
The only difference between the two models lies in the pre-training objective, as shown in \cref{sec:pretrain_loss}.

\paragraph{Pre-training Dataset.}
For foundation model pre-training, we curate a large-scale human kidney scRNA-seq corpus from the CELLxGENE Census \cite{10.1093/nar/gkae1142} (version \texttt{2023-05-15}).
We filter datasets by Organism (\textit{Homo sapiens}) and \emph{Tissue} (kidney), and restrict to RNA modality.
This yields a heterogeneous collection of kidney cells spanning multiple studies, donors, and experimental conditions, with all constituent datasets listed in \cref{sec:pretrain-detail}.
In total, the pre-training corpus contains \textbf{800K} cells.
\paragraph{Pre-training Recipe.}
We train Cell-JEPA using the pre-training objective described in \cref{sec:pretrain_loss}. For the scGPT baseline, we follow the training objective described in \cref{sec:scgpt} and \citet{Cui2024}. Implementation details and hyperparameter settings for training the Cell-JEPA model are provided in \cref{sec:pretrain-detail}.

\subsection{Task 1: Finetuning on PBMC}
\label{subsec:task1}

\paragraph{Task Description.}
We extract cell embeddings from both Cell-JEPA and scGPT and evaluate representation quality primarily through quantitative clustering metrics, complemented by qualitative UMAP visualizations.
We report biologically motivated evaluation metrics following \citet{Cui2024, Luecken2022}.

\paragraph{Dataset.}
We evaluate on a peripheral blood mononuclear cell (PBMC) dataset, PBMC-10k, consisting of two scRNA-seq batches obtained from a healthy donor (nominally 4k and 8k cells) \citep{10x_pbmc4k,10x_pbmc8k}.
The data are preprocessed following the \texttt{scvi-tools} PBMC pipeline \cite{Gayoso2022}.
The processed dataset contains 3,346 highly variable genes, with 7,982 cells in the first batch and 4,008 cells in the second batch \cite{Gayoso2022}.
Cell-type annotations cover major immune populations, including B cells, CD4+ and CD8+ T cells, NK cells, monocytes, dendritic cells, and others.

\paragraph{Training/Validation Split.}
We adopt the data splitting strategy from \cite{Cui2024}.
We randomly split the PBMC 10K dataset into training and validation sets with a 9:1 ratio, using a fixed random seed of 42 to ensure reproducibility. UMAP visualizations were generated using the full PBMC 10K dataset.

\paragraph{Finetuning Recipe.}
We finetune our pre-trained Cell-JEPA and scGPT checkpoints on each dataset using the finetuning objectives described in \cref{sec:finetune-loss}. The implementation detail and hyperparameter
we use for finetuning the Cell-JEPA model can be found in \cref{sec:finetune-detail}.

\paragraph{Evaluation Metrics.}
We consider four embedding-quality metrics: average biological conservation (AvgBIO), normalized mutual information ($\mathrm{NMI}_{\mathrm{cell}}$), average silhouette width ($\mathrm{ASW}_{\mathrm{cell}}$), and adjusted Rand index ($\mathrm{ARI}_{\mathrm{cell}}$).
We provide detailed definitions in \cref{def:AvgBIO}.

\begin{figure*}[tb!]
    \centering
    \begin{subfigure}[t]{0.24\textwidth}
        \centering
        \includegraphics[width=\linewidth,trim=0 2cm 0 0,clip]{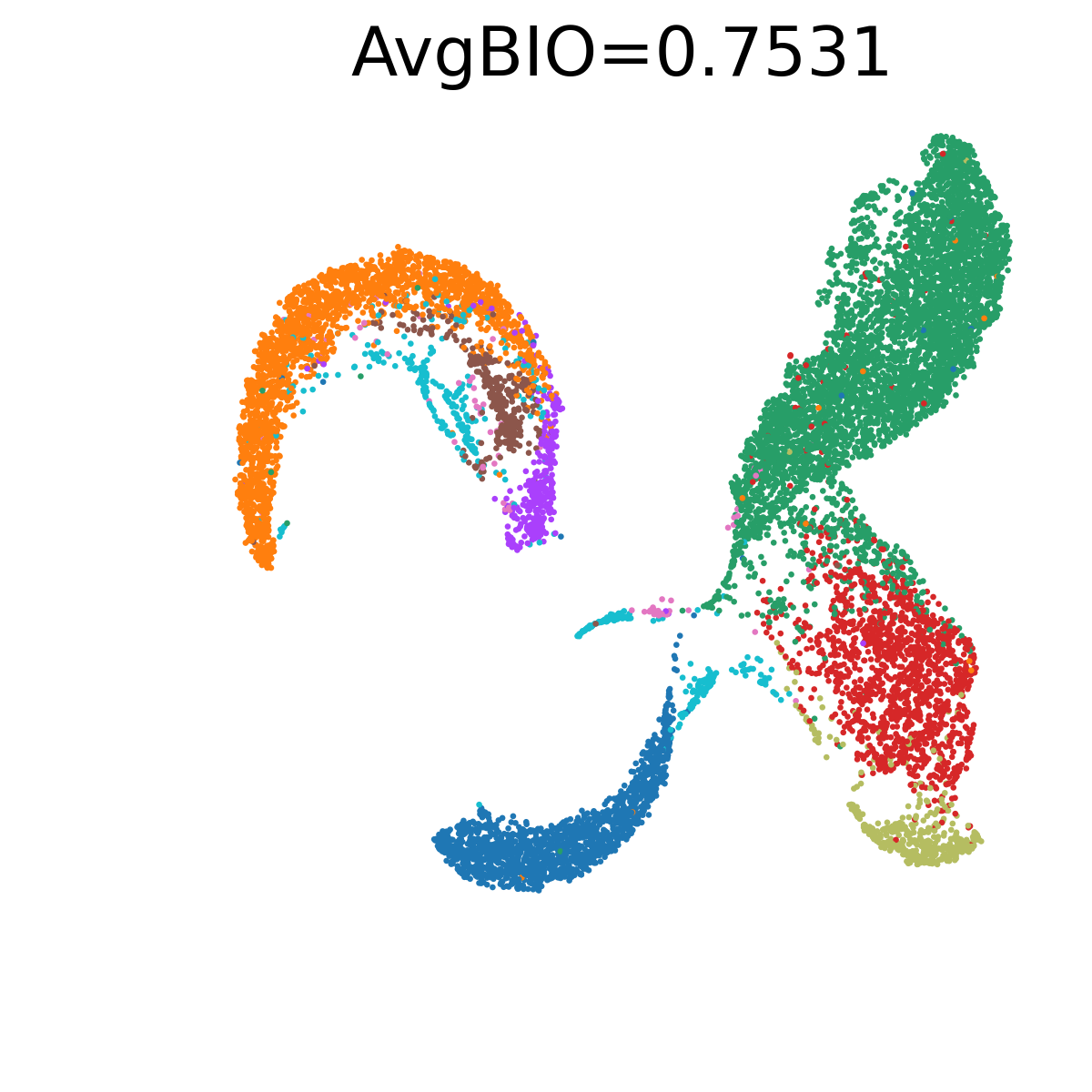}

        \caption{scGPT with finetuning}
        \label{fig:umap-original}
    \end{subfigure}
    \hfill
    \begin{subfigure}[t]{0.24\textwidth}
        \centering
        \includegraphics[width=\linewidth,trim=0 2cm 0 0,clip]{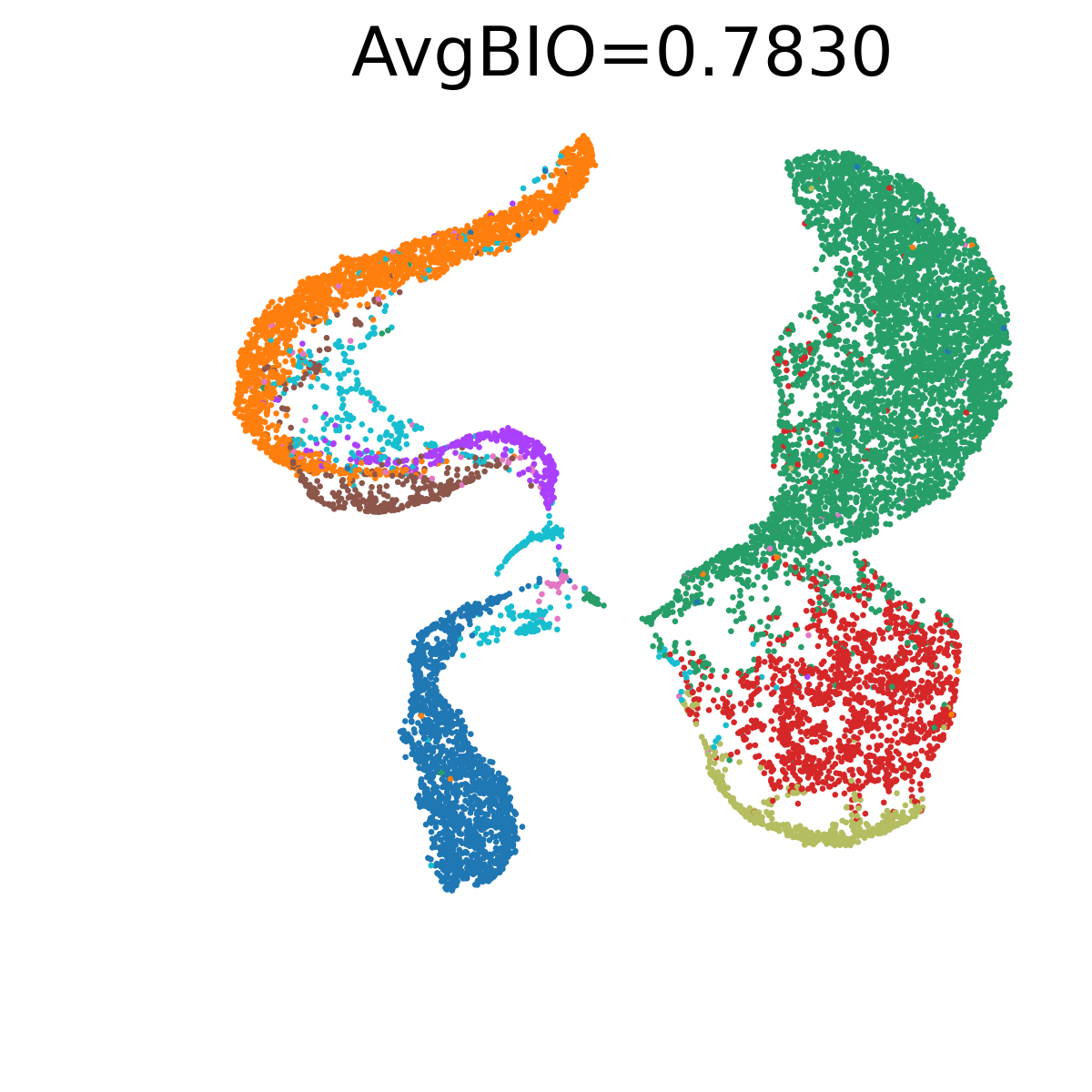}
        \caption{Cell-JEPA with finetuning}
        \label{fig:umap-jepa}
    \end{subfigure}
    \hfill
    \begin{subfigure}[t]{0.24\textwidth}
        \centering
        \includegraphics[width=\linewidth,trim=0 2cm 0 0,clip]{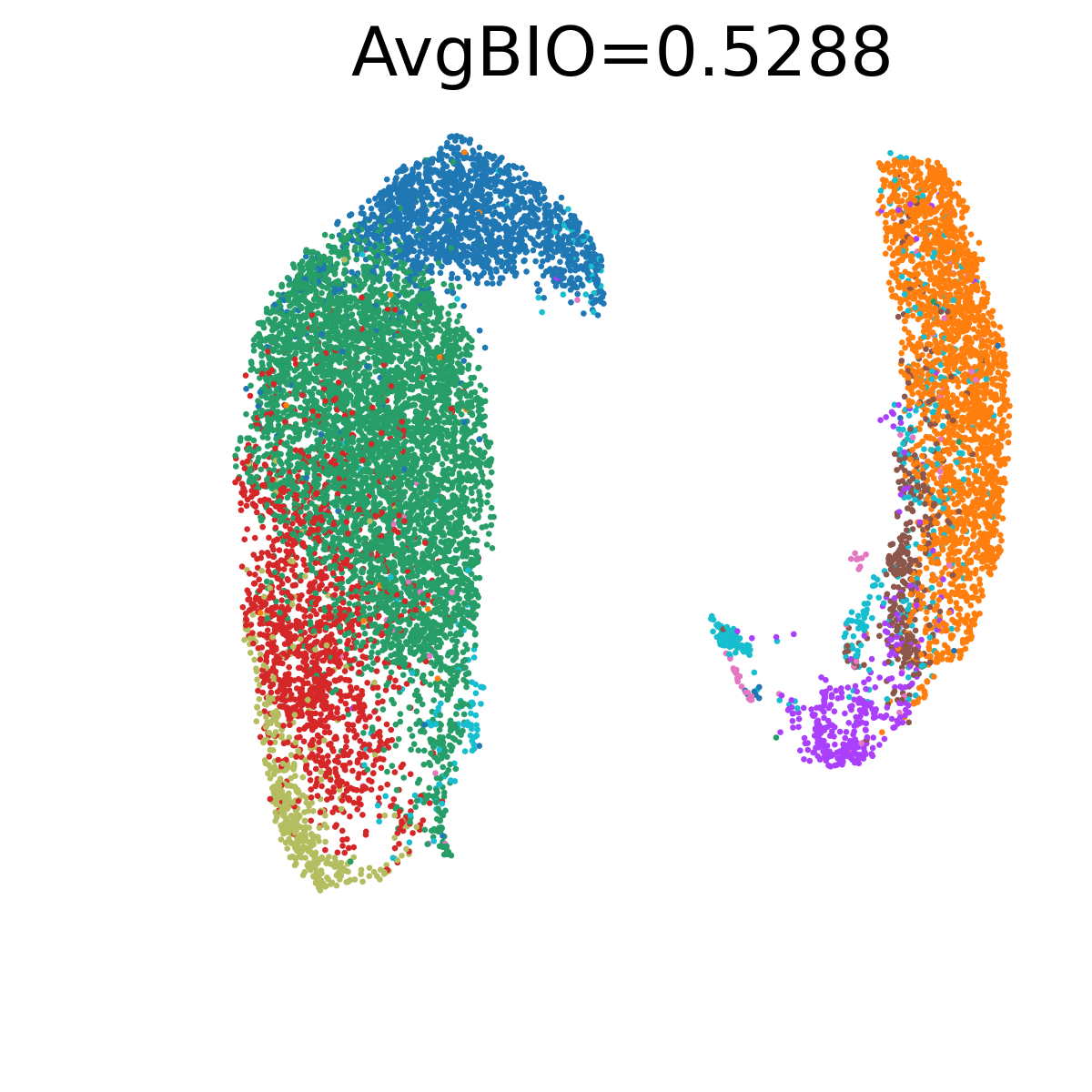}
        \caption{scGPT under zero-shot}
        \label{fig:umap-original-zero-shot}
    \end{subfigure}
    \hfill
    \begin{subfigure}[t]{0.24\textwidth}
        \centering
        \includegraphics[width=\linewidth,trim=0 2cm 0 0,clip]{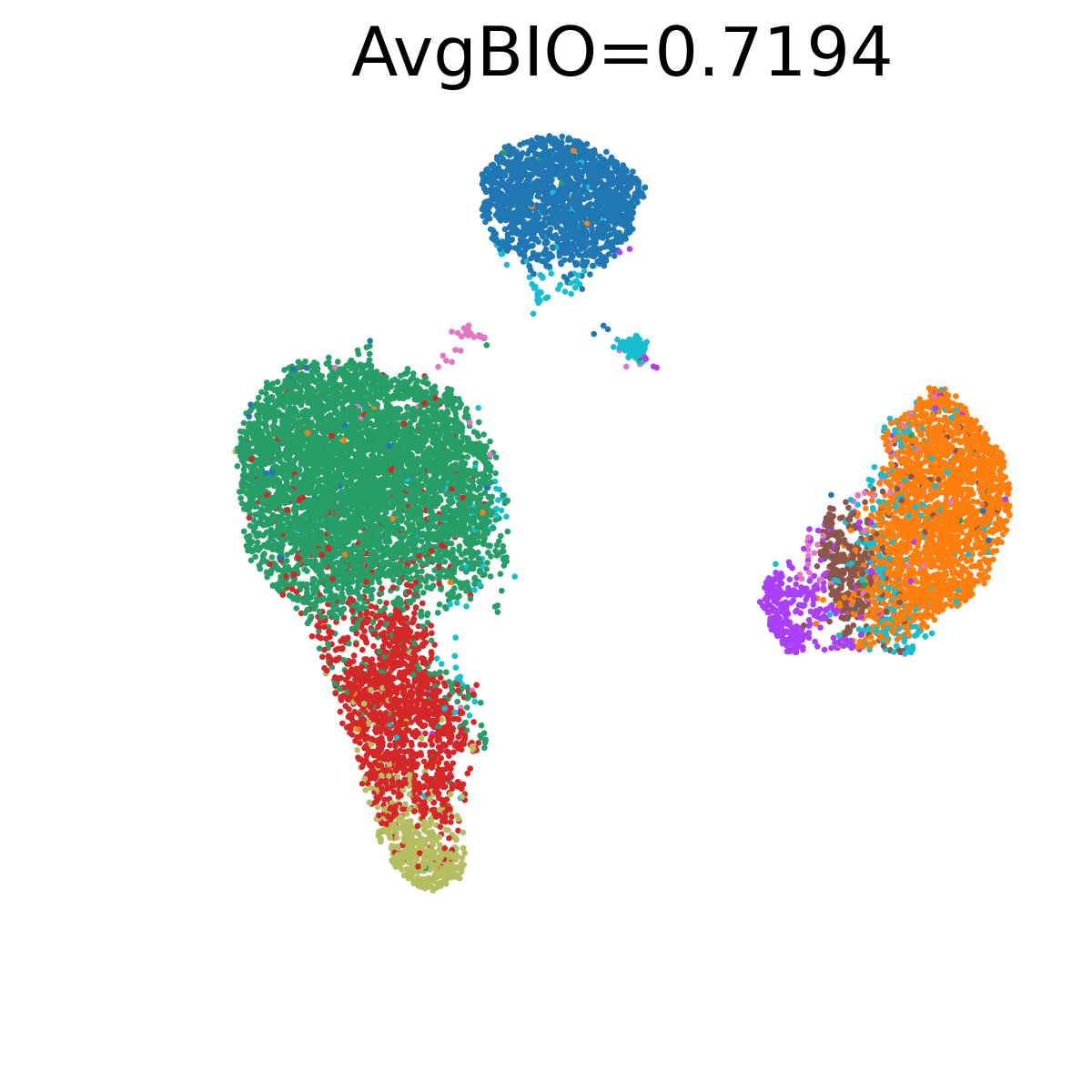}
        \caption{Cell-JEPA under zero-shot}
        \label{fig:umap-jepa-zero-shot}
    \end{subfigure}
    \caption{\textbf{Comparison of Cell Embeddings from scGPT and Cell-JEPA.}
UMAP visualizations of cell embeddings on the PBMC-10K.
(a) scGPT after finetuning (Task 1),
(b) Cell-JEPA after finetuning (Task 1),
(c) scGPT under zero-shot evaluation (Task 2), and
(d) Cell-JEPA under zero-shot evaluation (Task 2).}
\end{figure*}

\paragraph{Results.}
We generate single-cell embeddings for PBMC-10k and visualize them using UMAP, showing that the learned latent spaces of both models exhibit qualitative separation of annotated cell types. \cref{fig:umap-original,fig:umap-jepa}.
Compared with the baseline scGPT model, Cell-JEPA achieves improved performance across all four clustering metrics (\cref{tab:finetuned}).
These results suggest that augmenting scGPT with the JEPA objective facilitates learning more informative representations from single-cell gene expression data.

\begin{table}[htbp]
\centering

\begin{minipage}[t]{0.48\textwidth}
\centering
\caption{
\textbf{Performance of Task 1: Finetuning on PBMC.}
AvgBIO score, $\mathrm{NMI}_{\mathrm{cell}}$, $\mathrm{ASW}_{\mathrm{cell}}$, and $\mathrm{ARI}_{\mathrm{cell}}$ metrics for Cell-JEPA and scGPT after finetuning on PBMC-10K dataset.
After task-specific finetuning, Cell-JEPA outperforms scGPT across all four metrics.}
\label{tab:finetuned}

\resizebox{\textwidth}{!}{%
\begin{tabular}{l|cccc}
\toprule
Model & AvgBIO & $\mathrm{NMI_{cell}}$ & $\mathrm{ASW_{cell}}$ & $\mathrm{ARI_{cell}}$ \\
\midrule
scGPT & 0.7531 & 0.7652 & 0.7100 & 0.7842 \\
Cell-JEPA & \textbf{0.7830} & \textbf{0.7761} & \textbf{0.7256} & \textbf{0.8472} \\
\bottomrule
\end{tabular}
}
\end{minipage}
\hfill
\begin{minipage}[t]{0.48\textwidth}
\centering
\caption{
\textbf{Performance of Task 2: Zero-shot Evaluation on PBMC.}
AvgBIO score and $\mathrm{NMI}_{\mathrm{cell}}$, $\mathrm{ASW}_{\mathrm{cell}}$, and $\mathrm{ARI}_{\mathrm{cell}}$ metrics for pre-trained Cell-JEPA and scGPT model under zero-shot evaluation on the PBMC-10K dataset.
Cell-JEPA outperforms scGPT.}
\label{tab:zero-shot}

\resizebox{\textwidth}{!}{%
\begin{tabular}{l|cccc}
\toprule
Model & AvgBIO & $\mathrm{NMI_{cell}}$ & $\mathrm{ASW_{cell}}$ & $\mathrm{ARI_{cell}}$ \\
\midrule
scGPT & 0.5288 & 0.5585 & 0.5329 & 0.4951 \\
Cell-JEPA & \textbf{0.7194} & \textbf{0.7690} & \textbf{0.5811} & \textbf{0.8081} \\
\bottomrule
\end{tabular}
}
\end{minipage}

\end{table}

\subsection{Task 2: Zero-shot on PBMC}
\label{subsec:task2}

Next, we evaluate zero-shot representation quality without any task-specific finetuning.
We perform inference using the pre-trained checkpoints and evaluate embeddings using the same metrics as in \cref{subsec:task1}.

\paragraph{Results.}

We generate single-cell embeddings for PBMC-10K via zero-shot inference and visualize them using UMAP (\cref{fig:umap-original-zero-shot,fig:umap-jepa-zero-shot}), which suggests stronger zero-shot separation of annotated cell types for Cell-JEPA than for scGPT.
Moreover, across quantitative metrics, Cell-JEPA consistently outperforms scGPT in the zero-shot setting; for instance, it improves AvgBIO by $36\%$ relative to scGPT (\cref{tab:zero-shot}).
Overall, these results indicate that the JEPA objective improves transferability of learned representations, enabling strong performance without downstream finetuning.

\subsection{Task 3: Perturbation Prediction}
\label{subsec:perturbation}

\paragraph{Task Description.}

We explore whether improved representations benefit perturbation-response 
prediction within a fixed biological context. Given a cell's baseline 
expression and a genetic perturbation, the goal is to predict post-perturbation 
expression. Both evaluation datasets (Norman, Adamson) use the K562 cell line; 
we therefore assess within-context generalization to held-out perturbations 
rather than cross-context transfer.

\paragraph{Dataset.}
We evaluate perturbation-response prediction on two Perturb-seq CRISPR screening datasets profiled in the K562 leukemia cell line: Adamson \cite{Adamson2016} and Norman \cite{Norman2019}.
Adamson is a CRISPRi Perturb-seq dataset containing 86 single-gene perturbations (GEARS-processed), together with matched control cells. Norman is a CRISPRa Perturb-seq dataset containing 105 single-gene perturbations and 131 two-gene perturbations (GEARS-processed), with hundreds of cells per perturbation on average.
All datasets are processed using the GEARS pipeline \cite{roohani2023predicting}.

\paragraph{Training/Validation/Testing Split.}
We follow the GEARS \texttt{simulation} split protocol, which partitions data at the perturbation level rather than the cell level.
Under this setting, perturbations in the test set are held out during training.
In Norman, held-out two-gene perturbations are further grouped into \texttt{combo\_seen2}, \texttt{combo\_seen1}, and \texttt{combo\_seen0}, depending on whether both, one, or neither constituent gene was observed as a single-gene perturbation during training \cite{roohani2023predicting}.

\paragraph{Finetuning Recipe.}
We finetune \emph{scGPT} and our \emph{Cell-JEPA} model using the official scGPT perturbation finetuning pipeline, which leverages GEARS for dataset preprocessing, split construction, and evaluation. In this setup, a Transformer-based predictor (scGPT, also adapted for Cell-JEPA) is trained to model post-perturbation gene expression.
Each finetuning data point consists of (i) an unperturbed (control) gene expression profile, (ii) binary perturbation indicators specifying the perturbed gene(s), and (iii) the ground-truth post-perturbation expression profile.
Both models are trained for 15 epochs with learning rate $10^{-4}$ and batch size 64.
The training takes \texttt{2} hours on a single \texttt{A100} GPU. We provide the full training configuration details in \cref{sec:perturb-detail}.

\paragraph{Evaluation Metrics.}

Following GEARS \cite{roohani2023predicting}, we evaluate perturbation prediction using Pearson correlation between predicted and observed post-perturbation gene expression vectors. We report \texttt{pearson} over all genes, \texttt{pearson\_de} restricted to differentially expressed genes, and \texttt{top20\_de\_non\_dropout} restricted to top-ranked DE genes after filtering dropout genes.
Full details are provided in \cref{app_subsec:task3_eval}.

\begin{table}[htbp]
\centering

\begin{minipage}[t]{0.48\textwidth}
\centering
\caption{
\textbf{Performance of Task 3: Perturbation Prediction on Unseen Single-gene Perturbations.}
Compared with scGPT, Cell-JEPA improves absolute post-perturbation state prediction and DE-focused recovery on both the Norman \cite{Norman2019} and Adamson \cite{Adamson2016} datasets.
\texttt{pear} denotes \texttt{pearson}, \texttt{pear\_de} denotes \texttt{pearson\_de}, and \texttt{top20} denotes \texttt{top20\_de\_non\_dropout}.
}
\label{tab:pert-main}

\resizebox{\textwidth}{!}{%
\begin{tabular}{llccc}
\toprule
Dataset & Model & \texttt{pear} & \texttt{pear\_de} & \texttt{top20} \\
\midrule
Norman  & scGPT      & 0.631 & 0.276 & 0.278 \\
Norman  & Cell-JEPA  & \textbf{0.787} & \textbf{0.592} & \textbf{0.565} \\
\midrule
Adamson & scGPT      & 0.905 & 0.699 & 0.677 \\
Adamson & Cell-JEPA  & \textbf{0.937} & \textbf{0.741} & \textbf{0.720} \\
\bottomrule
\end{tabular}
}
\end{minipage}
\hfill
\begin{minipage}[t]{0.48\textwidth}
\centering
\caption{
\textbf{Delta-based Metrics on Unseen Single-gene Perturbations.}
Delta correlations remain low for both models, and Cell-JEPA shows reduced performance on both datasets, with the largest decrease observed on Adamson \citep{Adamson2016}.
\texttt{$\Delta$\_pear} denotes \texttt{pearson\_delta}, \texttt{$\Delta$\_pear\_de} denotes \texttt{pearson\_delta\_de}, and \texttt{$\Delta$\_top20} denotes \texttt{pearson\_delta\_top20\_de\_non\_dropout}.
}
\label{tab:pert-delta}

\resizebox{\textwidth}{!}{%
\setlength{\tabcolsep}{4.5pt}
\begin{tabular}{llccc}
\toprule
Dataset & Model & \texttt{$\Delta$\_pear} & \texttt{$\Delta$\_pear\_de} & \texttt{$\Delta$\_top20} \\
\midrule
Norman  & scGPT & \textbf{0.103} & \textbf{0.105} & \textbf{0.169} \\
Norman  & JEPA  & 0.059 & 0.041 & 0.115 \\
\midrule
Adamson & scGPT & \textbf{0.173} & \textbf{0.101} & \textbf{0.120} \\
Adamson & JEPA  & 0.120 & $-0.176$ & $-0.152$ \\
\bottomrule
\end{tabular}
}
\end{minipage}

\end{table}

\paragraph{Results.}~
\label{sec:pert-result}
\cref{tab:pert-main} summarizes performance on GEARS \texttt{unseen\_single} perturbations for Adamson and Norman.
Cell-JEPA consistently improves both absolute state prediction and DE-focused recovery relative to the scGPT baseline, with the largest gains observed on Norman.
These results suggest that augmenting scGPT finetuning with a latent-space JEPA objective yields representations that better generalize to unseen genetic perturbations.

We also report delta-based metrics (\texttt{pearson\_delta}, \texttt{pearson\_de\_delta}, \texttt{delta\_top20\_de\_non\_dropout}), which evaluate correlation on expression changes relative to the control state (post-perturbation minus control), and thus emphasize effect-size accuracy rather than absolute post-perturbation state prediction.
As shown in \cref{tab:pert-delta}, both scGPT and Cell-JEPA achieve low delta correlations on unseen single-gene perturbations, indicating that accurate change-from-control estimation remains challenging in this regime.
While Cell-JEPA improves absolute-state metrics (\cref{tab:pert-main}), it does not consistently improve delta-based metrics and exhibits the largest decrease on Adamson.

A plausible explanation is that the JEPA objective encourages learning features 
invariant across masked views, which benefits recognizing cellular states but 
may suppress sensitivity to perturbation-induced deviations. This suggests 
that representation learning (what state is this cell in?) and perturbation 
modeling (how does a cell change?) address complementary aspects of cellular 
prediction and may require distinct approaches. Overall, we view these results as highlighting a trade-off between absolute-state fidelity and effect-size accuracy, and suggest that scaling to larger perturbation datasets and/or explicitly training with delta-aware objectives may be necessary to fully realize the benefits of latent-space predictive learning for perturbation generalization.

\section{Conclusion}
\label{sec:conclusion}
We introduced Cell-JEPA, a joint-embedding predictive framework that 
complements reconstruction-based pre-training with latent-space prediction. 
By predicting cell-level embeddings rather than reconstructing noisy counts, 
Cell-JEPA learns representations robust to the dropout artifacts that 
dominate single-cell measurements.

Our primary finding is a 36\% relative improvement in zero-shot cell-type 
clustering, demonstrating that latent prediction yields representations 
that transfer without task-specific finetuning. On perturbation prediction 
within a single cell line, we observe improved absolute-state reconstruction 
but not effect-size estimation. This suggests that learning robust cell 
representations and modeling perturbation effects are complementary challenges 
that may require distinct approaches.

These results establish latent-space prediction as a promising direction for 
single-cell representation learning, while highlighting that perturbation 
modeling remains an open problem likely requiring explicit context modeling 
or greater contextual diversity in training data. We leave limitations in \cref{appsec:limit} and related work in \cref{sec:related}.

\section*{Acknowledgments}
We thank the Quest high-performance computing facility at Northwestern University 
for computational resources and staff support; Quest is jointly supported by 
the Office of the Provost, the Office for Research, and Northwestern University 
Information Technology.

H.L.\ was supported by NIH R01LM1372201, NSF AST-2421845, Simons Foundation 
MPS-AI-00010513, AbbVie, Dolby, and a Chan Zuckerberg Biohub Chicago Spoke Award. 
A.A.K.\ was supported by NIH DP2AI177884, a Chan Zuckerberg Investigator Award, 
and a Common Mechanisms of Autoimmunity Insight Award. 
M.K.\ was supported by NIH T32GM139782.

The content is solely the responsibility of the authors and does not necessarily 
represent the official views of the funding agencies.

\def\arxivfont{\rm}
\bibliographystyle{plainnat}

\bibliography{refs}

\clearpage
\appendix
\label{sec:append}
\part*{Appendix}
{
\setlength{\parskip}{-0em}
\startcontents[sections]
\printcontents[sections]{ }{1}{}
}

{
\setlength{\parskip}{-0em}
\startcontents[sections]
\printcontents[sections]{ }{1}{}
}

\section{Limitations}
\label{appsec:limit}
Despite the improvements demonstrated by Cell-JEPA, five limitations remain.
\begin{itemize}
    \item \textbf{Limited Pre-training Dataset.}
    Due to computational constraints, we pre-train our models on a kidney-specific dataset rather than the full human single-cell atlas used by \citet{Cui2024}, which contains approximately 33 million cells.
    From a biological perspective, Cell-JEPA is therefore exposed to a limited diversity of cell types, and its ability to generalize to other tissues, rare cell populations, and disease states has not been systematically evaluated.
    To ensure a fair comparison with the scGPT baseline under this constraint, we adopt identical hyperparameters and mini-batch sizes, which preclude task-specific hyperparameter optimization for Cell-JEPA.

    \item \textbf{Model Scale.}
    We constrain the model size to 28 million parameters (single student or teacher model), which may be insufficient to fully leverage substantially larger and more heterogeneous pre-training datasets.
    Scaling Cell-JEPA to larger model capacities and conducting large-scale foundation model pre-training are left to future work.

    \item \textbf{Limited Evaluation Scope across Model Families.}
    Our experimental evaluation applies the JEPA training framework exclusively to scGPT and compares performance with and without JEPA under this fixed backbone.
    We do not extend the JEPA objective to other single-cell foundation model architectures, and assessing its generality across alternative training pipelines remains an important direction for future exploration.

    \item \textbf{Delta-based Evaluation Metrics.}
    While our method improves absolute post-perturbation state prediction, performance on delta-based evaluation metrics remains limited.
    This suggests persistent challenges in accurately modeling effect-size changes relative to control conditions.

    \item \textbf{Residual Technical and Biological Confounders.}
    Despite the robustness of latent-space JEPA objectives, learned embeddings may still reflect residual batch effects, donor-specific variability, and differences in sequencing depth, which could influence downstream biological interpretation.

    \item \textbf{Perturbation evaluation scope.}
    Our perturbation experiments use two datasets from a single cell line (K562), 
    assessing generalization to held-out perturbations within a fixed biological 
    context. We do not evaluate cross-context transfer (e.g., across cell types, 
    donors, or activation states), which recent work suggests is the primary 
    challenge for perturbation prediction \cite {ahlmann2025deep}. We also do not 
    compare against simple baselines such as mean or additive models, which have 
    proven competitive in systematic evaluations \cite{vinas2025systema}. Our 
    perturbation results should therefore be interpreted as preliminary exploration 
    rather than a solution to the perturbation prediction problem.

\end{itemize}

\section{Related Work}
\label{sec:related}

In this section, we review prior work on the development of single-cell foundation models and the application of joint embedding predictive architectures in other domains.

\paragraph{Single Cell Foundation Models.}
Large foundation models have demonstrated strong potential for modeling complex data across diverse domains, including language and biology \cite{zhou2025genomeocean, wu2025genome, zhou2024dnabert, brown2020language}.
The adaptation of large foundation models to transcriptomic data has reshaped representation learning in single-cell biology.
Early deep probabilistic approaches such as scVI \cite{Lopez2018} introduced variational inference with encoder--decoder architectures to model scRNA-seq counts, learning latent representations that help account for batch effects and library-size variation.
Building on this foundation, Transformer-based methods learn context-aware representations by treating each cell as a collection of gene tokens (often arranged into a pseudo-sequence) and applying self-supervised objectives.
Within this paradigm, masked modeling approaches such as scBERT \cite{scBERT} and Geneformer \cite{Theodoris2023} pre-train Transformer encoders to recover masked genes from cellular context, while generative methods such as scGPT \cite{Cui2024} and TranscriptFormer \cite{Pearce2025} learn to model gene expression by predicting missing values under structured conditioning schemes.
Recent work has focused on scaling model capacity and incorporating cross-species knowledge.
For instance, scFoundation \cite{Hao2024} leverages xTrimoGene to scale pre-training to over 50 million human cells \cite{Hao2024}, while GeneCompass \cite{Yang2024} introduces a knowledge-informed cross-species framework trained on over 120 million transcriptomes \cite{Yang2024}.
Similarly, Universal Cell Embeddings (UCE) \cite{Rosen2023} utilizes protein language models to construct a unified latent space for cells that generalizes across tissues and species.

A remaining challenge is that many current pre-training objectives are closely tied to \emph{gene-level reconstruction} in the observed measurement space---predicting masked counts, masked gene identities, or related entry-wise targets.
Because scRNA-seq measurements are sparse and strongly affected by technical variation (e.g., capture efficiency and sampling noise), optimizing such objectives can encourage representations that preserve nuisance variation in addition to biology \cite{lahnemann2020eleven}.
This motivates complementary objectives that enforce invariances directly in representation space.
In contrast, our approach learns single-cell representations through latent-space prediction, aiming to reduce sensitivity to superficial measurement noise while better capturing stable, high-level cellular programs.

\paragraph{Joint Embedding Predictive Architecture.} 
Joint Embedding Architectures (JEA) provide a unifying framework for self-supervised representation learning methods. 
In JEAs, the model encodes and projects multiple views of the same data point into a shared latent space, and their training objectives encourage consistency across views. 
A central challenge in this setting is representation collapse, where the encoder converges to producing identical representations regardless of the input.
\textbf{J}oint \textbf{E}mbedding \textbf{P}redictive \textbf{A}rchitectures (\textbf{JEPA}) \cite{lecun_2022} address limitations of direct alignment objectives by shifting the learning signal from representation matching to latent-space prediction. 
Instead of enforcing similarity between embeddings of different views, JEPA trains a predictor to map the representation from a context view to the representation of a target view. 
In common instantiations, a separate target encoder produces the target representation and is updated using an exponential moving average of the context encoder weights, while the prediction loss is computed entirely in representation space rather than input space.
This predictive formulation eliminates the need for negative samples and avoids explicit reconstruction of observed tokens or values, which can otherwise bias learning toward low-level details.
JEPA has shown success across multiple domains, including I-JEPA \cite{assran2023selfsupervisedlearningimagesjointembedding} for images, V-JEPA \cite{assran2025vjepa2selfsupervisedvideo} for videos, and LLM-JEPA \cite{huang2025llmjepalargelanguagemodels} for language modeling.
It demonstrates strong performance and improved scalability. 
By focusing on predicting embeddings rather than reconstructing raw measurements, JEPA promotes learning meaningful and compositional features.

In our setting, gene expression values reflect complex regulatory programs and cell-state--specific signals. 
Effective modeling requires representations to capture both gene-level variation and cell-level semantics.
GeneJEPA \cite{Litman2025} is a recent JEPA-style foundation model for transcriptomics that applies latent-space prediction to masked gene subsets, supporting strong transfer across downstream tasks. 
While conceptually aligned with our goal of moving beyond pure reconstruction-based pre-training, our approach differs in a key way: Cell-JEPA combines a JEPA representation-level objective with an explicit gene-level reconstruction loss.
We find that retaining a reconstruction anchor helps preserve fine-grained gene-expression information and stabilizes optimization, while JEPA-style prediction encourages invariances and higher-level structure in the learned representations.
This hybrid objective enables Cell-JEPA to learn cell embeddings that are both robust to measurement noise and informative for downstream cellular prediction tasks.

\section{scGPT}
\label{sec:scgpt}
\label{sec:subsection_A}

Unlike language, gene expression vectors are inherently \emph{unordered} (genes do not form a natural sequence), so scGPT \cite{Cui2024} introduces a specialized attention mask and iterative generation scheme that is autoregressive over \emph{sets of genes}, rather than over a fixed gene order.

\subsection{Tokenization and Input Representation}
Each cell is represented as a set of gene tokens with associated expression values.
The transformer input typically consists of: (i) a dedicated \texttt{<cls>} token that yields a global cell embedding, (ii) a subset of \emph{known} genes with observed expression values, and (iii) a subset of \emph{unknown} genes whose expression values are masked and must be predicted.
Gene identities are embedded using a learnable gene embedding table, while expression values are embedded through a value embedding module (and optionally additional condition tokens such as batch or perturbation labels).

\subsection{Autoregressive Gene Generation via Masked Attention}
To enable generation without imposing an arbitrary gene ordering, scGPT uses a structured attention mask that enforces a \emph{known} $\rightarrow$ \emph{unknown} dependency.
When predicting an unknown gene, its token is allowed to attend to all known genes (and itself), but it cannot attend to other unknown genes.
Let $\mathcal{U}_{\mathrm{unk}}$ denote indices of unknown genes in the current step. The attention mask is defined as:
\begin{align}
a_{i,j} =
\begin{cases}
0, & \text{if } j \notin \mathcal{U}_{\mathrm{unk}},\\
0, & \text{if } i=j \text{ and } j\in \mathcal{U}_{\mathrm{unk}},\\
-\infty, & \text{if } i\neq j \text{ and } j\in \mathcal{U}_{\mathrm{unk}}.
\end{cases}
\label{eq:scgpt_mask}
\end{align}
This prevents information leakage between simultaneously masked targets while allowing each unknown gene to condition on the full known context.

\subsection{Pretraining Objective}
Given final hidden states $\{h_j\}$ from the transformer, scGPT predicts expression values of unknown genes via an MLP head and optimizes mean squared error over masked positions:
\begin{align*}
\mathcal{L}_{\mathrm{GEP}}
=
\frac{1}{|\mathcal{U}_{\mathrm{unk}}|}
\sum_{j\in \mathcal{U}_{\mathrm{unk}}}
\left(\mathrm{MLP}(h_j) - v_j \right)^2,
\end{align*}
where $v_j$ is the ground-truth expression of gene $j$.

In addition to \emph{gene-prompt} prediction (predict masked genes from observed genes), scGPT also describes a \emph{cell-prompt} generation mode where the global cell embedding (from \texttt{<cls>}) is used as a prompt to generate remaining gene expression values \cite{Cui2024}. In practice, these modes can be applied sequentially and their losses combined during pretraining.

\subsection{Iterative Autoregressive Inference}
At inference time, scGPT performs generation in $K$ iterative rounds, yielding an autoregressive procedure over gene subsets rather than a fixed gene order.
At each iteration $t$, the model computes predictions for the remaining unknown genes using the mask in \eqref{eq:scgpt_mask}.
It then \emph{selects} a subset of the most confident predictions and promotes them to the known set for the next iteration.
Concretely, scGPT uses a confidence-based expansion strategy where approximately a $1/K$ fraction of genes with highest confidence are added each round until all genes are generated \cite{Cui2024}.
This creates an autoregressive refinement process: high-confidence genes are generated earlier and subsequently serve as context for harder-to-predict genes in later rounds.

\subsection{Issues with the ECS Loss in scGPT}
\label{sec:scgpt-ecs}
The original scGPT framework \cite{Cui2024} adopts an alternative formulation of the ECS loss:
\begin{align*}
\mathrm{ECS}_{\text{scGPT}} = \big(\cos{\mathbf{e}^i, \mathbf{e}^j} - \beta \big)^2,
\end{align*}
where $\mathbf{e}^i$ and $\mathbf{e}^j$ denote the cell embeddings of cells $i$ and $j$, respectively, and $\beta$ is a fixed similarity threshold set to $0.3$. While this formulation explicitly enforces pairwise cosine similarities toward a predefined target, it can inadvertently distort biologically meaningful relationships by artificially pushing distinct cellular states closer together or farther apart based solely on a manually chosen hyperparameter. In contrast, our method avoids reliance on such fixed thresholds, leading to a more robust preservation of intrinsic biological structure in the learned representation space.

\clearpage
\section{Evaluation Metrics}
We provide the detailed evaluation metrics calculation methods here.

\subsection{Average Biological Conservation (AvgBIO)}
\label{def:AvgBIO}
We adopt the AvgBIO score from \cite{Cui2024} to evaluate the quality of single-cell embeddings. The AvgBIO score is the average of three cell-type clustering metrics developed by \cite{Luecken2022}, including normalized mutual information ($\mathrm{NMI_{cell}}$), adjusted rand index ($\mathrm{ARI_{cell}}$), and
average silhouette width ($\mathrm{ASW_{cell}}$). In particular, the AvgBIO score is defined as follows:
\begin{align*}
\mathrm{AvgBIO}=\frac{\mathrm{NMI_{cell}}+\mathrm{ARI_{cell}}+\mathrm{ASW_{cell}}}{3}
\end{align*}
\paragraph{Normalized Mutual Information $\mathrm{NMI_{cell}}$.}
To quantify the agreement between ground-truth cell type annotations and clustering results derived from integrated cell embeddings, we compute the normalized mutual information (NMI) score. Louvain clustering is performed over a range of resolutions from $0.1$ to $2.0$ in increments of $0.1$, and the maximum NMI score across resolutions is reported. The resulting metric, referred to as $\mathrm{NMI_{cell}}$, measures the shared information between the predicted cluster assignments and true cell type labels. $\mathrm{NMI_{cell}}$ ranges from $0$ to $1$, with higher values indicating stronger correspondence between clusters and annotated cell types.

\paragraph{Adjusted Rand Index $\mathrm{ARI_{cell}}$.}
In addition to NMI, we employ the adjusted rand index (ARI) to assess the consistency between annotated cell type labels and the Louvain clusters optimized for NMI. Unlike the standard rand index, ARI corrects for agreements that occur by chance, providing a more robust measure of clustering accuracy. The ARI score for cell types, denoted as $\mathrm{ARI_{cell}}$, ranges from 0 to 1, where 0 corresponds to random label agreement and 1 indicates a perfect match between predicted clusters and ground-truth annotations.

\paragraph{Average Silhouette Width $\mathrm{ASW_{cell}}$.}
To evaluate the quality of cell-type separation in the integrated embedding space, we compute the average silhouette width (ASW). The silhouette width of each cell measures the relative similarity between its within-cluster distance and the distance to the nearest neighboring cluster, thereby capturing both cluster compactness and separation. By averaging silhouette widths across all cells, we obtain the ASW score for cell types, denoted as $\mathrm{ASW_{cell}}$. This score ranges from $-1$ to $1$, where higher values indicate well-separated and cohesive clusters, while values near or below zero suggest overlapping clusters or potential misclassification.

\subsection{Evaluation Metrics for Perturbation Prediction}
\label{app_subsec:task3_eval}

Following GEARS \cite{roohani2023predicting}, we evaluate perturbation-response prediction by comparing predicted and observed post-perturbation gene expression profiles using Pearson correlation. For each perturbation $p$ in the evaluation set, let $x^{(p)} \in \mathbb{R}^{G}$ denote the observed post-perturbation gene expression vector over $G$ genes. And let $\hat{x}^{(p)} \in \mathbb{R}^{G}$ denote the corresponding model prediction. The GEARS evaluation suite reports several Pearson correlation metrics computed on different gene subsets. Pearson correlation is computed over: all genes (global state matches), differentially expressed (DE) genes (perturbation-specific effects), and the top 20 DE genes with dropout-prone genes removed. 

For any two vectors $a,b \in \mathbb{R}^d$, Pearson correlation is:
\begin{align*}
\rho(a,b)
=
\frac{\sum_{i=1}^{d}(a_i-\bar{a})(b_i-\bar{b})}
{\sqrt{\sum_{i=1}^{d}(a_i-\bar{a})^2}\sqrt{\sum_{i=1}^{d}(b_i-\bar{b})^2}},
\end{align*}
where $\bar{a}$ and $\bar{b}$ are the component-wise means.
Unless otherwise stated, the reported metrics are aggregated across perturbations in the corresponding GEARS test subset (i.e. \texttt{unseen\_single}) by averaging $\rho(\hat{x}^{(p)}, x^{(p)})$ over $p$.

\paragraph{All-gene Correlation \texttt{pearson}.}

\texttt{pearson} computes $\rho(\hat{x}^{(p)}, x^{(p)})$ using all genes:
\begin{align*}
\texttt{pearson}
=
\mathbb{E}_{p}\big[\rho(\hat{x}^{(p)}, x^{(p)})\big].
\end{align*}
This metric emphasizes agreement with the absolute post-perturbation transcriptional state.

\paragraph{DE-restricted Correlation \texttt{pearson\_de}.}
For each perturbation $p$, a set of differentially expressed (DE) genes $\mathcal{D}_p$ was identified by comparing the perturbed and control conditions \cite{roohani2023predicting}.
\texttt{pearson\_de} computes Pearson correlation restricted to this DE set:
\begin{align*}
\texttt{pearson\_de}
=
\mathbb{E}_{p}\big[\rho(\hat{x}^{(p)}_{\mathcal{D}_p}, x^{(p)}_{\mathcal{D}_p})\big],
\end{align*}
where $x_{\mathcal{D}_p}$ denotes the subvector of $x$ indexed by $\mathcal{D}_p$.
This metric focuses evaluation on genes that exhibit a transcriptional response to the perturbation.

\paragraph{Top DE Genes with Non-dropout Filtering \texttt{top20\_de\_non\_dropout}.}
GEARS provides a filtered subset of top-ranked DE genes with dropout-prone genes removed.
Let $\mathcal{T}_p \subseteq \mathcal{D}_p$ denote the resulting set (typically $|\mathcal{T}_p|=20$) provided by the GEARS preprocessing \cite{roohani2023predicting}.
\texttt{top20\_de\_non\_dropout} computes correlation on this set:
\begin{align*}
\texttt{top20\_de\_non\_dropout}
=
\mathbb{E}_{p}\big[\rho(\hat{x}^{(p)}_{\mathcal{T}_p}, x^{(p)}_{\mathcal{T}_p})\big].
\end{align*}

\paragraph{Delta Variants (Change-from-control Correlation).}
GEARS also reports delta-based metrics that evaluate predicted \emph{changes relative to control}.
Let $x^{(\mathrm{ctrl})}\in\mathbb{R}^{G}$ denote the control expression reference (as provided by the GEARS pipeline \cite{roohani2023predicting}).
We define the observed and predicted deltas for perturbation $p$ as:
\begin{align*}
\Delta x^{(p)} = x^{(p)} - x^{(\mathrm{ctrl})},
\qquad
\Delta \hat{x}^{(p)} = \hat{x}^{(p)} - x^{(\mathrm{ctrl})}.
\end{align*}
The delta metrics compute the same correlations as above, but on $\Delta \hat{x}^{(p)}$ versus $\Delta x^{(p)}$:
\begin{align*}
\texttt{pearson\_delta}
&=
\mathbb{E}_{p}\big[\rho(\Delta\hat{x}^{(p)}, \Delta x^{(p)})\big],\\
\texttt{pearson\_de\_delta}
&=
\mathbb{E}_{p}\big[\rho(\Delta\hat{x}^{(p)}_{\mathcal{D}_p}, \Delta x^{(p)}_{\mathcal{D}_p})\big],\\
\texttt{delta\_top20\_de\_non\_dropout}
&=
\mathbb{E}_{p}\big[\rho(\Delta\hat{x}^{(p)}_{\mathcal{T}_p}, \Delta x^{(p)}_{\mathcal{T}_p})\big].
\end{align*}
These metrics emphasize delta (change-from-control) estimation and are sensitive to effect-size accuracy relative to the control baseline.

\clearpage
\section{Training Recipe}
We provide the detailed hyperparameter settings for pre-training and finetuning here.

\subsection{Pretraining Recipe}
\label{sec:pretrain-detail}
The pretrained model uses an embedding dimension of $512$ and comprises $12$ stacked transformer blocks, each with 8 attention heads and a hidden dimension of 512. 
The embedding alignment MLP consists of two fully connected layers with a GELU activation between them. We optimize the model using AdamW with a mini-batch size of $32$, a token-masking ratio of $0.15$, and an initial learning rate of $1\times10^{-4}$. We apply a weight decay of $2\times10^{-4}$ and decay the learning rate by a factor of $0.9$ after each epoch. We apply dropout with a probability of $0.2$ between each layer in the transformer head and the embedding alignment MLP.  Training is conducted on the Kidney dataset for a total of $4$ epochs. To account for differences in numerical scale, we assign a weight of $1000$ to the JEPA loss term and a weight of $1$ to the reconstruction loss term. The full training process takes approximately one day on a single \texttt{A5000} GPU.

We would like to acknowledge and thank the following authors whose datasets are used in the training of our foundation models and preparation of this paper: \cite{HCA_Kidney_UMich,The_Tabula_Sapiens_Consortium__2022,Wilson_2022,Suo_2022,Young_2021,Lake_2021,Zhang_2021,Muto_2021,Han_2020,Cao_2020,Stewart_2019,Young_2018}. 

\subsection{PBMC-10K Finetuning Recipe}
\label{sec:finetune-detail}
We finetune the model from a pretrained checkpoint trained on the Kidney dataset, as described in \cref{sec:pretrain-detail}. We train the model on the PBMC-10K dataset using a masking ratio of $40\%$. We inherit the JEPA loss weight from pretraining and set the weights of all other loss terms to $1$. We optimize the model using the Adam optimizer with a mini-batch size of $64$ and an initial learning rate of $1\times10^{-4}$, decaying the learning rate by a factor of $0.9$ after each epoch. We train the model for a total of $30$ epochs, and the finetuning process takes approximately $2$ hours on a single NVIDIA RTX A5000 GPU.

\subsection{Perturbation Finetuning Recipe} 
\label{sec:perturb-detail} 
We finetune the model for perturbation prediction using the Adamson and Norman datasets. The model is initialized from a checkpoint pretrained on the Kidney dataset. For the JEPA-enhanced model, we initialize the projector and predictor heads with random weights (or inferred from a checkpoint if available) and train them alongside the student encoder. We optimize using Adam with a learning rate of $1\times10^{-4}$ and a batch size of $64$. The learning rate is decayed by a factor of $0.9$ every epoch. Unlike the pretraining phase, we do not use masking on the input genes for perturbation prediction (mask ratio set to $0.0$). For the JEPA configuration, the student model is trained to predict the representations of a frozen teacher model (a copy of the student). The JEPA loss weight is set to $1.0$ (as to prevent representation drift due to out-of-distribution cell lines), and the reconstruction loss weight is set to $1.0$.
The contrastive ECS loss is activated and weighted at $0.8$. Training is conducted for $15$ epochs, which takes approximately $2$ hours on a single NVIDIA RTX A100 GPU. The model performance is evaluated using the GEARS evaluation suite (\cref{app_subsec:task3_eval}) on a held-out test set.

\end{document}